\documentclass{aastex}
\usepackage{spr-astr-addons}
\usepackage{url}
\RequirePackage{color}

\newcommand{\emaila}{authors@email.com}

\begin{document}
\title{The correlations of multi-wave band luminosity and BLR luminosity in Fermi 2LAC blazars}
\slugcomment{Not to appear in Nonlearned J., 45.}
\shorttitle{Short article title}
\shortauthors{Wang et al.}
\author{Zerui Wang}\and\author{Rui Xue}\and\author{Zhaohua Xie$^{\dag}$}\and\author{Leiming Du}\and\author{Tingfeng Yi}\and\author{Yunbing Xu}\and\author{Wenguang Liu}
\affil{Department of Physics, Yunnan Normal University, Kunming 650500, China\\$^{\dag}$email:zhxie007@126.com}
\email{\emaila}

\begin{abstract}
We collect a sample of 78 Fermi detected blazars with broad line region (BLR) data and the quasi-simultaneous multi-wave band data. By analyzing the sample, we find: (1) For whole blazar sample, there exist significant correlations between radio, $\gamma$-ray luminosity and BLR luminosity. The slope of the best-fit linear regression equation is close to the theoretical value 1. Our results provided another support of jet-disk symbiosis. (2) For FSRQ sample, we find a significant correlation between radio luminosity and optical luminosity, and weaker correlations between X-ray luminosity and radio, optical luminosity. It indicates that the X-ray band radiation mechanism is slightly different from radio and optical bands. The X-ray emission is diluted by inverse Compton scattering (IC). There is a significant correlation between BLR luminosity and $\gamma$-ray luminosity, but no correlations between X-ray luminosity and BLR, $\gamma$-ray luminosity, which suggests that the seed soft photons are different for IC process. From these results we suggest that the X-ray band emission is diluted by synchrotron self-Compton (SSC) process, $\gamma$-ray band emission is produced by external Compton (EC) process.
\end{abstract}
\section{Introduction}
Active galactic nuclei (AGNs) are the object which has high luminosity and their energy is believed come from the material which is captured by the gravity of the accretion by a supermassive black hole (Lynden - Bell, 1969). Blazars are the most extreme subclass of AGNs and characterized that highly relativistic jets directed toward our line of sight (Begelman et al., 1984; Urry $\&$ Padovani, 1995). Blazars are divided into flat spectrum radio quasars (FSRQs) and BL Lacs, FSRQs have strong emission lines(Equivalent Width, $EW>5 \mathring{A}$) while BL Lacs have very weak or no emission lines(EW$\leq 5 \mathring{A}$).

The typical spectral energy distribution (SED) is one of the important tools to help us study blazars. It consists of non-thermal radiation from jet and thermal radiation from accretion disk, BLR, torus and host galaxies. The thermal radiation can clearly see an independent bulge in ultraviolet from SED (blue bump). In addition to obvious thermal radiation, the typical SED of blazar is generally described by a double-bump structure. In the lepton models, the low energy peak is generally considered synchrotron radiation from relativistic electrons movement in the magnetic field, the range usually from radio band to ultraviolet or soft X-ray band. High energy peak is generally considered IC process, in which the high-energy electrons up-scatters the low energy photons (soft photons), the range usually from hard X-ray band to MeV band, even to TeV band. IC process is mainly divided into two kinds, one is SSC process, the soft photons are derived from the same area of synchrotron radiation (Konigl 1981; Marscher $\&$ Gear 1985; Ghisellini $\&$ Maraschi1989); the other one is EC process, the soft photons are derived from external photon fields, for example: accretion disk (Dermer $\&$ Schlickeiser 1993; Bloom $\&$ Marscher 1996; Bottcher, Mause $\&$ Schlickeiser 1997;), BLR (Sikora et al., 1994; Ghisellini $\&$ Madau 1996;), torus (Bla$\dot{z}$ejowski et al., 2000; Ghisellini $\&$ Tavecchio 2008) or the Cosmic Microwave Background (CMB, Bottcher et al., 2008; Yan et al., 2012; Meyer et al., 2015).

It is well known that the formation and acceleration processes of highly relativistic jets in AGN is one of the most fundamental open problems in astrophysics. There are two main explanations: 1. the origin of the jet kinetic energy is from Kerr black holes (BZ mechanism; Blandford $\&$ Znajek 1977); 2. the energy is come from the large-scale fields threading the rotating accretion disk (BP mechanism; Blandford $\&$ Payne, 1982). In both processes, the magnetic field must play a major role, so we can expect a relation between the accretion power and the jet power. (Maraschi $\&$ Tavecchio 2003).

The luminosity of the BLR can be taken as an indicator of the accretion power of the objects (Celotti et al. 1997). Czerny et al. (2004) analyzed the consequences of the hypothesis that the formation of BLR is intrinsically connected to the existence of cold accretion disks (Xie et al. 2004, 2006, 2007). The emission of blazars from radio to $\gamma$-ray band is dominated by the beamed non-thermal continuum radiation produced in the jet (Urry $\&$ Padovani 1995). So we can study the physics of relativistic jets and accretion disks by analyzing the correlation between the BLR emission and radio, optical, X-ray, $\gamma$-ray emissions (Rawlings $\&$ Saunders, 1991; Celotti et al. 1997; Cao $\&$ Jiang 1999, Xu et al. 1999; Xie et al. 2007). Serjeant et al. (1998) found a significant correlation between radio and optical band emission for a sample of steep-spectrum quasars, the result indicates a direct evidence of a close link between accretion onto BHs and the fueling of relativistic jets. The correlation between the luminosity in radio band and BLR for a sample of radio-loud quasars also was found by Cao $\&$ Jiang (1999).

By studying the correlation between luminosity in the radio, optical, X-ray and $\gamma$-ray bands from jet, we can find the restrictions of radiation mechanisms of different bands. Padovani et al. (1993), Stecker et al. (1993) and Dondi $\&$ Ghisellini (1995) found the existent correlation between the $\gamma$-ray luminosity and the radio luminosity of blazars. It indicates that $\gamma$-ray and radio emissions are produced by the same relativistic electrons. However, Muecke et al. (1997) and Fan et al.(1998) found the $\gamma$-ray emission is no correlation with 5GHz radio emissions. There is a positive correlation between between $F_{\gamma}$ and $F_{rad}$ (Cheng et al. 2000) while the present work states that there is no correlation between them (Cheng et al. 2000). Cheng et al. (2000) considered that the region of radio emission is possibly different from $\gamma$-ray emission. Xie et al. (1997) found that the correlation between the $\gamma$-ray luminosity and infrared luminosity is better than the correlation between the $\gamma$-ray luminosity and the luminosity in the optical or X-ray band. They suggested that the $\gamma$-ray emission is probably produced from IC process and the soft photons are derived from torus. In all of above cases, they didn't use simultaneous multi-wave bands data for the sources, but the simultaneous data is very critical for emission origin (Cheng et al. 2000).

Other researchers given the restrictions of radiation mechanisms of different bands by other ways. Studies of multiwavelength observation of blazars found that the synchrotron emission of FSRQs seems to extend only to soft X-ray band; the hard X-ray band derive from inverse Compton emission, probably from the SSC process(Catanese et al. 1997; Kubo et al. 1998; Mukherjee et al. 1999; Kataoka et al. 1999; Takahashi, Madejski $\&$ Kubo 1999). Inoue (1996) and Blazejowski (2000) succeed in fitting the X-band data with the SSC model. The produced region of $\gamma$-ray emission is also a current topic. Tavecchio et al. (2011) successfully used a two-zone model fitting the SED of PKS 1222+216, the VHE $\gamma$-ray emission is derived from a compact blob outside the BLR in their model.  Aleksic et al. (2011) found it can explain the SED of the Very high energy (VHE) $\gamma$-rays of PKS 1222+21, if the $\gamma$-ray emission is not produced within BLR. And many others considered  that the  locations of $\gamma$-ray emission region is outside BLR (Donea $\&$ Protheroe 2003; Liu $\&$ Bai 2006; Coogan 2016). However, Tavecchio and Ghisellini (2012) considered the VHE $\gamma$-ray emission produced within the BLR for FSRQs. Bottcher $\&$ Els (2016) confirmed that the $\gamma$-ray emission region is inside the inner boundary of the BLR.

In this parer, the (quasi-) simultaneous data are used to study the correlation	 between the luminosity in the BLR and radio, optical, X-ray, $\gamma$-ray bands.  We provide a new support in observation of jet-disk symbiosis and try to find the radiation mechanisms of different bands. In section 2 we present our whole blazar sample detected by Fermi $\gamma$-ray Space Telescope; in section 3 we give the method of correlation analysis and results; the section 4 is discussions and conclusions. 
\section{Sample}
In Xue et al.(2016), they collect a sample which contains 200 FSRQs and 79 BL Lacs from the second LAT AGN catalog(2LAC). At least one of the two components (synchrotron and IC components) of the blazars in their sample can be fitted by sufficient multifrequency data coverage. They construct the SEDs of all the blazars from the (quasi-) simultaneous multi-frequency data by using the ASDC SED Builder (It is an on-line service developed at the ASI Science Data Center. The database of the ASDC SED Builder provides observational data from many space telescopes and ground-based telescopes; Stratta 
et al. 2011). They use the log-parabolic law:
\begin{equation}
\log(\nu F_{\nu})=c(\log{\nu}) ^ {2} + b(\log{\nu}) + a
\end{equation}
to fit the synchrotron component and IC component separately. The luminosities and frequencies (in the rest frame) can be calculated through $(\nu L_{\nu})_{syn,IC} = 4\pi D_{L}^{2}(\nu F_{\nu})_{syn,IC}$ and $\nu_{syn,IC} = (1 +z)\nu^{obs}_{syn,IC}$, where $D_{L}$ is the luminosity distance and z is the redshift. During fitting the SEDs, the thermal component that prominent in the UV waveband has been identified and excluded by visual inspection. In this paper we are going to study the correlation of multi-wave band luminosity and BLR luminosity, therefore the simultaneous multiwavelength observations are important. The data of 2LAC catalog are collected over the first 24 months of the mission from 2008 August to 2010 August. In Xue et al. (2016), we find that the observation date of the simultaneous data of radio, optical and X-ray bands are more likely to be simultaneous with the 2LAC observations. Therefore, we have deliberately not used a more recent 3LAC catalog based on 4 years of Fermi/LAT data (2008 August - 2012 August) (Ackermann et al. 2015). Ultimately, 63 FSRQs and 15 BL Lacs with the complete SEDs and the data of BLR luminosity in Xue et al. (2016) are taken as our sample.

According to Abdo et al. (2010), blazars are classified as low synchrotron peak sources(LSPs) if $\log \nu_{p}(Hz) \leq 14.0$, intermediate synchrotron peak sources (ISPs), if $14.0 < \log \nu_{p}(Hz) \leq 15.0$, high synchrotron peak sources (HSPs) if $\log \nu_{p}(Hz) > 15.0$. Our sample is dominated by LSPs.

The relevant data for 78 blazars are listed in Table 1. The columns in this table are as follows.

1. Fermi catalogue name;

2. Other name;

3. Type of the Blazar (F = flat-spectrum radio quasar; B = BL Lac object; L = LSPs; I = ISPs; H = HSPs);

4. Redshift;

5. The observation date of the simultaneous data;

6. The logarithm of the 5GHz radio luminosity in units of  erg$\cdot$ s$^{-1}$;

7. The logarithm of the optical luminosity at rest wavelength $\sim$ 5100 $\mathring{A}$ in units of erg$\cdot$ s$^{-1}$;

8. The logarithm of the X-ray luminosity about 1keV in units of erg$\cdot$ s$^{-1}$;

9. The logarithm of the $\gamma$-ray luminosity bo 100MeV in units of erg$\cdot$ s$^{-1}$;

10. The logarithm of the BLR luminosity in units of erg$\cdot$ s$^{-1}$;

11. References for BLR data: 1: Xue et al. (2016); 2: Xiong $\&$ Zhang (2014); 3: Sbarrato (2012); 4: Celotti (1997); 5: Cao $\&$ Jiang (2016); 6: Xie $\&$ Li (2012).

\section{Correlation analysis and results}
Using the fitting parameters of the second-degree polynomial function, we calculate the monochromatic luminosity for each band (radio band corresponding to 5GHz, optical band corresponding to 5100 $\mathring{A}$, X band corresponding to 1keV and $\gamma$ band corresponding to 100MeV). Then we analyze the correlation between the BLR luminosity and radio, optical, X-ray, $\gamma$-ray luminosity, moreover correlations of the luminosity between each band (radio, optical, X-ray, $\gamma$-ray band) from jet also have been considered. However, there may exist bias to the correlation results from the effect of strong correlation between the luminosity and  redshift, So we should use the partial correlation to avoid the  redshift effect. If $r_{ij}$ is the correlation coefficient between $r_{i}$ and $r_{j}$ , in the case of three variables, the correlation between two of them, excluding the effect of the third one, is
\begin{equation}
r_{12,3}=\frac{r_{12}-r_{13}r_{23}}{\sqrt{1-r_{13}^2}\sqrt{1-r_{23}^2}}.
\end{equation}

The results of partial correlation for whole sample are shown in table 2, and for FSRQ sample are shown in table 3. Besides, according to data of Table 1, we use linear regression analysis to study the correlation of the luminosity of the BLR with radio, optical, X-ray, $\gamma$-ray band. The results are shown in Table 4.

In addition, we find that the correlation between each band has the following features:

1. There are significant correlation between $L_{BLR}$ and $L_{rad}$, $L_{\gamma}$ in the whole sample and the FSRQ sample;

2. The correlation between $L_{rad}$ and $L_{opt}$ is strong in the FSRQ sample;

3. The correlation between $L_{X}$ and $L_{rad}$, $L_{opt}$ is strong in the FSRQ sample, but weaker than the correlation between $L_{rad}$ and $L_{opt}$;

4. There is no correlation between $L_{X}$ and $L_{\gamma}$ in the FSRQ sample.

5. There is no correlation between $L_{X}$ and $L_{BLR}$ in the FSRQ sample.

6.There are significant correlations between $\gamma$-ray luminosity and the radio, optical luminosity in the FSRQ sample.

\section{Discussions and conclusions}

\subsection{jet and disk connection}

Ghisellini $\&$ Gabriele (2006) found that if relativistic jets are powered by a Poynting flux, under some reasonable assumption the power in BZ mechanism can be written as
\begin{equation}
L_{jet}\sim \frac{L_{disk}}{\eta},
\end{equation}
where $\eta$ is the conversion efficiency of mass into energy in an active nucleus.

In addition, based on the present theories of accretion disk the BLR is photo-ionized by nuclear source (probably radiation from the disk). Maraschi $\&$ Tavecchio (2003) obtained
\begin{equation}
L_{BLR}=\tau L_{disk},
\end{equation}
where $\tau$ is the fraction of the central emission reprocessed by the BLR, usually assumed to be 0.1.

From equations (3) and (4), we get (Xie $\&$ Zhang 2012)
\begin{equation}
L_{BLR}= \tau \eta L_{jet}.
\end{equation}

From equation (5), we have
\begin{equation}
\log L_{BLR}=\log L_{jet} + \log \eta + const.
\end{equation}
The equations (6) shows that: 1. the theoretical predicted coefficient of $\log L_{BH} - \log L_{jet}$ linear relation is 1; 2. $\eta$ is not the same for all objects, $\log \eta$ will contributes to the dispersion around the linear relation, but the impact is not very obvious(Xie et al. 2005).

The results in Table 2 show that there are correlations between BLR and the four bands(radio, optical, X-ray, $\gamma$-ray band), and the slope  of the linear regression equation is nearly 1 for whole blazar sample, the results are consistent with the theoretical prediction of equations (6). Especially, slopes of the linear relations between BLR luminosity and radio, $\gamma$-ray luminosity are respectively 0.999, 1.086 which are almost 1. Our results provide a solid experimental basis for the theory of Maraschi $\&$ Tavecchio (2003) and Ghisellini $\&$ Gabriele (2006).

The BLR emission can be taken as an indicator of the accretion power (Celotti et al. 1997), and the radio emission is a good indicator of the jet power (Blandford $\&$ Konigl 1979, Serjeant etal. 1998, Xie et al.2007). In addition, both the radio emission and $\gamma$-ray emission are produced in the jet (Urry $\&$ Padovani 1995). So the intrinsic correlation between $L_{BLR}$ and $L_{rad}$, $L_{\gamma}$ indicate that the formation of jet and accretion disk of Kerr black hole exist a close connection, in other words, the jet and accretion disk is couple and symbiosis for blazar, which is consistent with the result of Ghisellini et al (2009), Xie et al. (2008,2012). Our provide a new support in observation.

\subsection{Radiation mechanism of various wave bands from jet.}
FSRQs are the majority of our sample, so we just discuss the radiation mechanism of FSRQs.

There is a significant correlation between $L_{rad}$ and $L_{opt}$ for FSRQ sample, which implies that the emissions of radio and optical are produced by the synchrotron radiation in the jet.This result is consistents with the expectation of the lepton model (e.g., Dermer $\&$ Schlickeiser 1993; Sikora et al. 1994; Inoue $\&$ Takahara 1996; Katarzynski et al. 2001).

Our results in table 3 show significant correlations between $L_{X}$ and $L_{rad}$, $L_{opt}$, which are weaker than the correlation between $L_{rad}$ and $L_{opt}$. It indicates that the X-ray emission might come from a mixture of both SSC process and synchrotron radiation. Other researchers (e.g., Paliya et al. 2015; Kang 2017)  have also reached the same results.

As already mentioned in the introduction section, the possible origins of the soft photons include synchrotron radiation (SSC), accretion disk, BLR, torus and CMB. The fact that there exists a weak but still significant correlation between $L_{BLR}$ and $L_{\gamma}$ and this might be interpreted that the photons from the BLR are only one of the sources of the soft photons needed for the IC process. However, the SSC mechanism do not dominate the $L_{\gamma}$ and the details is as follows.  

All FSRQs are LSPs in our sample, so their synchrotron peak frequencies is less than $10^{14}$ Hz. If very extreme conditions are not taken into account, the SSC peak frequency rarely reaches the $\gamma$-ray frequency considered in table 3 ($10^{22.38}$ Hz) according to the common homogeneous single zone SSC model. Moreover, another evidence supporting our conclusion is that there is no correlation between $L_{X}$ and $L_{\gamma}$ (p=0.03), which is supposed to present if the SSC mechanism is the dominating process of the $L_{X}$. Some researchers (e.g., Ghisellini et al. 2002; Celotti $\&$ Ghisellini 2008; Ghisellini et al. 2010; Finke 2013)  have also obtained similar results in different ways. Such as Xue et al. (2016) have analyzed the slope of the best linear regression that given by the correlation between the IC luminoisty and the synchrotron luminosity, they suggested that the high-energy component of FSRQs is dominated by EC process.

In Fan et al. (2016), non-simultaneous multi-wavelength data were compiled for a sample of 1425 Fermi 3LAC blazars to calculate their SEDs. They studied correlations between gamma-ray emissions and lower energy monochromatic emissions. They found significant correlations between $\gamma$-ray luminosity and the radio, optical luminosity, a weak corrleation between $\gamma$-ray luminosity and X-ray luminosity. And they considered that the  $\gamma$-ray emission is strongly believed to come from EC process and the X-ray emission  is come from a mixture of both EC process and synchrotron radiation. The results are similar with us except that we think the X-ray emission is come from a mixture of both SSC process and synchrotron radiation.

At the same time, we have calculated the different soft photon energy density with radiation location changes(Figure 1). The energy density approximated by formulae for BLR and torus is come from Hayashida et al. (2012), for CMB is come from Ghisellini $\&$ Tavecchio (2009), for disk is come from Sikora et al. (2009). When the radiation zone is located between $r_{BLR}$ and $r_{IR}$, the energy density of BLR and torus are almost equal and both of them dominate the soft photons together. When the radiation region is located very close to the  central engine, the energy density of BLR and disk are almost equal. When the radiation region is far away from the central black hole (beyond 100pc), the BLR, torus and CMB should dominate the soft photons together.Therefore, at least for our sample, the candidates for the sources of the soft photons for $L_{\gamma}$ include  BLR, disk, torus, and CMB.

In summary, for most FSRQs, We consider the emissions of radio and optical are produced by the synchrotron radiation. The X-ray emission is come from the mixture of SSC and synchrotron process. The $\gamma$-ray emission is dominated by EC process and the possible sources are mixture of BLR and torus, of BLR and disk, or of BLR, torus and CMB.

\begin{acknowledgements}
This work is supported by the Joint Research Fund in Astronomy (Grant Nos U1431123, 11263006, 10978019, 11463001) under cooperative agreement between the National Natural Science Foundation of China (NSFC) and Chinese Academy of Sciences (CAS), the Provincial Natural Science Foundation of Yunnan (Grant No.: 2013FZ042) and the Yunnan province education department project (Grant No. 2014Y138, 2016ZZX066).
\end{acknowledgements}

\clearpage
\begin{figure}
\includegraphics[width=84mm]{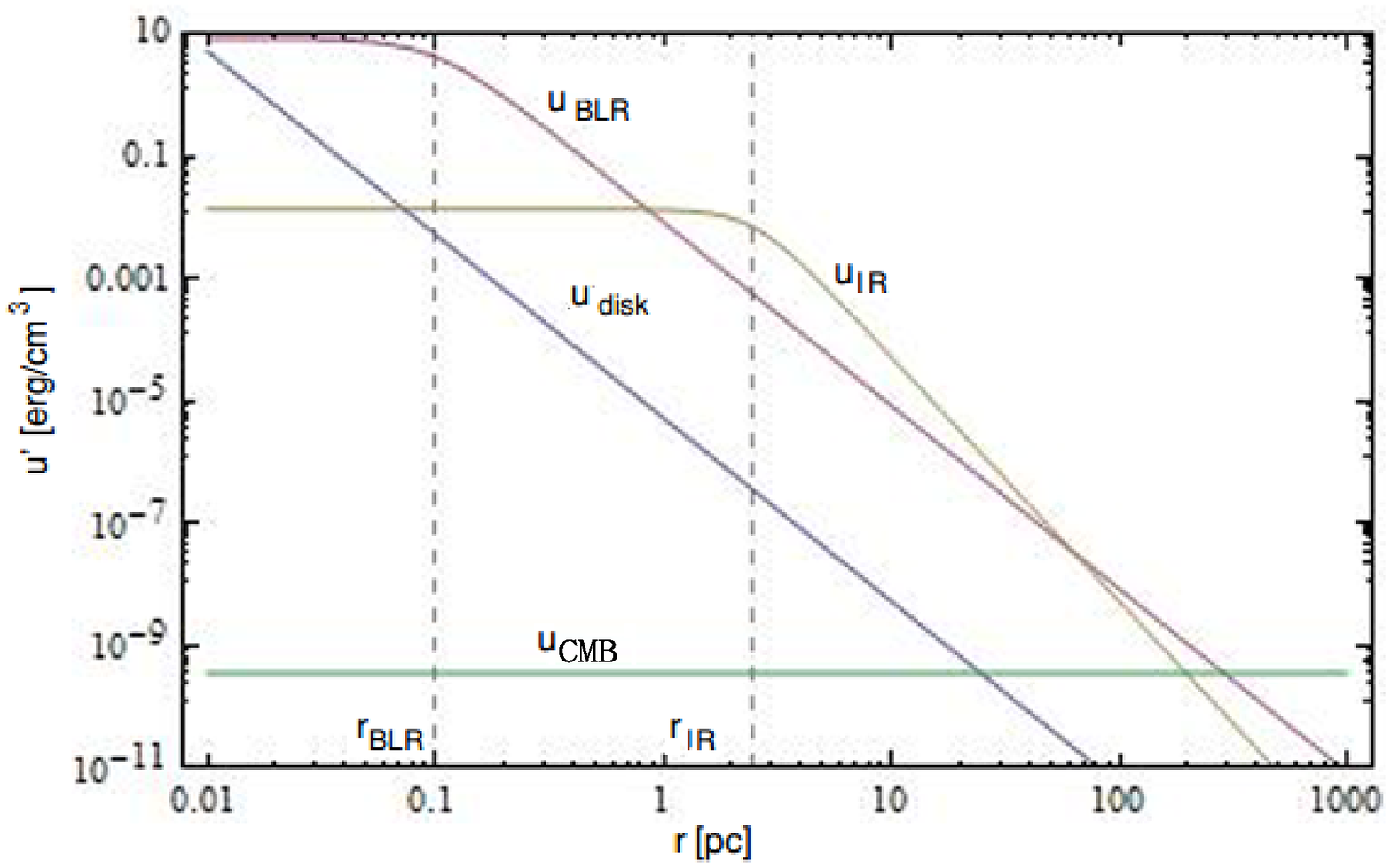}
\caption{The soft photon Energy densities of accretion disk (blue line), BLR (red line), dust (yellow line), and CMB (green line) as seen in the
jet comoving frame, as a function of distance from the central engine. $r_{BLR}$ and $r_{IR}$ are  characteristic radii of BLR and torus.}
\end{figure}

\clearpage
\begin{deluxetable}{crcccccccccccccrl}
\tabletypesize{\scriptsize} \rotate
\tablecaption{The sample\label{tbl-1}} \tablewidth{0pt}
\tablehead{\colhead{Fermi Name} & \colhead{other name} & \colhead{type} & \colhead{Redshift} & \colhead{time} & \colhead{$\log \nu L_{rad}$} & \colhead{$\log \nu L_{opt}$} & \colhead{$\log \nu L_{X}$}  & \colhead{$\log \nu L_{\gamma}$} & \colhead{$\log \nu L_{BLR}$} & \colhead{Ref.}\\
\colhead{(1)} & \colhead{(2)} & \colhead{(3)} & \colhead{(4)} &
\colhead{(5)} & \colhead{(6)} & \colhead{(7)} & \colhead{(8)} &
\colhead{(9)} & \colhead{(10)} & \colhead{(11)}}
\startdata
2FGLJ0237.8+2846   & 4C +28.07                   & F-L       & 1.213    & 2010/2/5                 & 44.99      & 46.79      & 44.94    & 47.82        & 45.90      & 1    \\
2FGL J0407.7+0740  & TXS 0404+075                & F-L       & 1.133    & 2010/2/14-2010/2/15      & 43.97      & 45.68      & 44.86    & 46.80        & 44.51      & 1    \\
2FGLJ0423.2-0120   & PKS 0420-01                 & F-L       & 0.916    & 2009/8/27                & 45.41      & 47.09      & 45.54    & 47.35        & 44.63      & 1    \\
2FGL J0439.0-1252  & PKS 0436-129                & F-L       & 1.285    & 2010/7/1                 & 44.30      & 45.80      & 45.01    & 46.73        & 44.78      & 1    \\
2FGL J0531.8-8324  & pks 0541-834                & F-L       & 0.774    & 2009/11/8-2009/11/11     & 43.27      & 45.07      & 44.43    & 46.51        & 43.74      & 2    \\
2FGL J0532.7+0733  & OG 505                      & F-L       & 1.254    & 2010/4/25                & 44.97      & 46.13      & 45.43    & 47.47        & 44.86      & 1    \\
2FGL J0601.1-7037  & PKS 0601-70                 & F-L       & 2.409    & 2010/3/9                 & 44.43      & 47.02      & 44.99    & 47.92        & 44.69      & 1    \\
2FGL J0608.0-0836  & PKS 0605-085                & F-L       & 0.872    & 2010/6/7                 & 44.82      & 45.43      & 42.57    & 46.65        & 44.97      & 1    \\
2FGL J0654.2+4514  & B3 0650+453                 & F-L       & 0.928    & 2010/3/23                & 43.99      & 45.70      & 44.82    & 46.94        & 44.26      & 1    \\
2FGL J0654.5+5043  & GB6 J0654+5042              & F-I       & 1.253    & 2010/1/15                & 43.92      & 46.46      & 45.05    & 46.85        & 43.97      & 1    \\
2FGL J0656.2-0320  & TXS 0653-033                & F-L       & 0.634    & 2010/3/31                & 43.98      & 44.76      & 40.70    & 46.55        & 45.68      & 1    \\
2FGL J0714.0+1933  & MG2 J071354+1934            & F-L       & 0.54     & 2010/4/2-2010/4/3        & 43.23      & 45.70      & 43.98    & 46.41        & 43.93      & 1    \\
2FGL J0746.6+2549  & B2 0743+25                  & F-L       & 2.979    & 2010/10/15               & 44.88      & 46.04      & 46.15    & 48.06        & 45.46      & 1    \\
2FGL J0750.6+1230  & PKS 0748+126                & F-L       & 0.889    & 210/4/1-2010/4/12        & 44.72      & 46.10      & 45.64    & 46.56        & 44.95      & 1    \\
2FGL J0805.5+6145  & TXS 0800+618                & F-L       & 3.033    & 2010/4/3-2010/4/4        & 45.36      & 46.29      & 46.01    & 47.93        & 45.56      & 1    \\
2FGL J 0824.9+5552 & OJ535                       & F-L       & 1.41812  & 2010/3/28-2010/4/8       & 44.55      & 45.80      & 45.48    & 47.07        & 45.31      & 1    \\
2FGL J0830.5+2407  & S3 0827+24                  & F-L       & 0.942    & 2010/4/18-2010/4/19      & 44.40      & 45.70      & 44.75    & 45.74        & 44.98      & 2    \\
2FGL J0909.1+0121  & PKS 0906+01                 & F-L       & 1.024    & 2010/4/18-2010/5/5       & 44.64      & 46.42      & 45.75    & 47.55        & 45.20      & 1    \\
2FGL J0912.1+4126  & B3 0908+41                  & F-L       & 2.563    & 2010/2/25                & 43.66      & 45.98      & 45.25    & 47.21        & 45.36      & 1    \\
2FGL J0920.9+4441  & B3 0917+449                 & F-L       & 2.19     & 2009/10/29               & 45.25      & 46.76      & 46.18    & 48.24        & 45.78      & 1    \\
2FGL J0956.9+2516  & B2 0954+25A                 & F-L       & 0.707    & 2010/5/7-2010/6/15       & 44.27      & 45.50      & 45.09    & 46.43        & 44.93      & 1    \\
2FGL J0957.7+5522  & 4C+55.17                    & F-L       & 0.896    & 2009/11/1                & 44.64      & 46.26      & 45.07    & 47.00        & 44.58      & 1    \\
2FGL J1037.5-2820  & PKS B1035-281               & F-L       & 1.066    & 2010/1/22-2010/1/23      & 44.04      & 45.97      & 44.62    & 46.91        & 44.95      & 1    \\
2FGL J1106.1+2814  & MG2 J110606+2812            & F-L       & 0.843    & 2010/5/24                & 43.68      & 45.92      & 44.51    & 46.57        & 45.16      & 1    \\
2FGL J1146.8-3812  & PKS 1144-379                & F-L       & 1.048    & 2010/6/24                & 44.67      & 46.14      & 45.53    & 46.82        & 44.48      & 1    \\
2FGL J1159.5+2914  & 4C+29.45                    & F-L       & 0.724    & 2010/5/28-2010/6/11      & 44.60      & 46.29      & 45.01    & 47.02        & 44.68      & 1    \\
2FGL J1206.0-2638  & PKS 1203-26                 & F-L       & 0.789    & 2010/6/30-2010/12/10     & 44.28      & 45.77      & 45.27    & 46.38        & 44.07      & 1    \\
2FGL J1208.8+5441  & CRATES J1208+5441           & F-L       & 1.344    & 2010/5/16-2010/5/21      & 44.18      & 46.18      & 45.19    & 47.43        & 44.52      & 1    \\
2FGL J1229.1+0202  & 3C 273                      & F-L       & 0.158    & 2010/6/7-2010/6/23       & 44.55      & 45.28      & 45.85    & 46.22        & 45.54      & 2    \\
2FGLJ1256.1-0547   & 3C 279                      & F-L       & 0.536    & 2010/1/15                & 45.29      & 46.00      & 45.42    & 47.24        & 44.78      & 1    \\
2FGLJ1310.6+3222   & 1Jy 1308+326                & F-L       & 0.997    & 2009/12/12-2009/12/21    & 45.26      & 46.16      & 45.32    & 47.17        & 44.96      & 1    \\
2FGL J1326.8+2210  & B2 1324+22                  & F-L       & 1.4      & 2010/3/30                & 44.73      & 45.99      & 45.38    & 47.39        & 44.96      & 1    \\
2FGL J1333.5+5058  & CLASS J1333+5057            & F-L       & 1.362    & 2010/3/13                & 43.56      & 45.70      & 44.43    & 46.94        & 44.37      & 1    \\
2FGL J1337.7-1257  & PKS 1335-127                & F-L       & 0.539    & 2010/1/18$\sim$2010/1/26 & 44.48      & 45.76      & 45.35    & 46.27        & 44.18      & 1    \\
2FGL J1347.7-3752  & PMN J1347-3750              & F-L       & 1.3      & 2010/9/20                & 43.91      & 46.22      & 45.16    & 47.15        & 44.67      & 1    \\
2FGL J1358.1+7644  & S5 1357+76                  & F-L       & 1.585    & 2010/5/24-2010/5/25      & 44.66      & 45.65      & 45.06    & 47.02        & 44.20      & 1    \\
2FGL J1408.8-0751  & PKS 1406-076                & F-L       & 1.494    & 2010/5/23                & 44.77      & 46.51      & 45.41    & 47.39        & 45.47      & 1    \\
2FGL J1436.9+2319  & PKS 1434+235                & F-L       & 1.548    & 2010/6/14                & 44.76      & 46.03      & 45.04    & 46.84        & 44.72      & 1    \\
2FGL J1504.3+1029  & PKS 1502+106                & F-L       & 1.839    & 2010/7/29                & 45.27      & 46.84      & 45.32    & 48.63        & 45.24      & 1    \\
2FGL J1514.6+4449  & BZQ J1514+4450              & F-L       & 0.57     & 2010/4/6                 & 42.65      & 44.89      & 43.85    & 45.95        & 43.33      & 1    \\
2FGL J1539.5+2747  & MG2 J153938+2744            & F-L       & 2.191    & 2010/3/17                & 44.41      & 46.52      & 45.47    & 46.98        & 44.63      & 1    \\
2FGL J1549.5+0237  & PKS 1546+027                & F-L       & 0.414    & 2010/2/13-2010/2/20      & 43.82      & 45.27      & 44.74    & 46.06        & 44.80      & 1    \\
2FGLJ1635.2+3810   & 4C 38.41                    & F-L       & 1.814    & 2010/3/7                 & 45.65      & 46.79      & 45.44    & 48.64        & 45.48      & 1    \\
2FGL J1637.7+4714  & 4C+47.44                    & F-L       & 0.735    & 2010/7/30-2010/7/31      & 44.12      & 45.43      & 45.15    & 46.62        & 44.58      & 1    \\
2FGL J1709.7+4319  & B3 1708+433                 & F-L       & 1.027    & 2009/12/1                & 43.73      & 45.97      & 45.22    & 47.15        & 44.03      & 1    \\
2FGL J1733.1-1307  & PKS 1730-130                & F-L       & 0.902    & 2010/3/14-2010/4/10      & 45.21      & 46.69      & 45.47    & 47.27        & 44.83      & 1    \\
2FGL J1848.5+3216  & B2 1846+32                  & F-L       & 0.798    & 2010/10/6-2010/10/19     & 44.02      & 45.97      & 45.56    & 46.78        & 44.58      & 1    \\
2FGL J1924.8-2912  & PKS B1921-293               & F-L       & 0.353    & 2010/9/30                & 44.57      & 45.60      & 45.17    & 46.09        & 44.02      & 1    \\
2FGL J1958.2-3848  & PKS 1954-388                & F-L       & 0.63     & 2010/4/9-2010/4/14       & 44.39      & 45.77      & 45.10    & 46.71        & 44.20      & 1    \\
2FGL J1959.1-4245  & PMN J1959-4246              & F-L       & 2.178    & 2010/4/5-2010/4/14       & 44.28      & 46.67      & 45.40    & 47.89        & 45.13      & 1    \\
2FGL J2135.6-4959  & PMN J2135-5006              & F-L       & 2.181    & 2010/4/22-2010/5/5       & 44.42      & 45.74      & 45.24    & 47.52        & 45.26      & 1    \\
2FGL J2144.8-3356  & PMN J2145-3357              & F-L       & 1.361    & 2009/9/22-2009/9/24      & 43.80      & 46.31      & 44.71    & 46.97        & 44.18      & 1    \\
2FGL J2157.4+3129  & B2 2155+31                  & F-L       & 1.488    & 2009/7/8-2009/7/12       & 44.57      & 46.12      & 45.19    & 47.39        & 44.74      & 1    \\
2FGL J2201.9-8335  & PKS 2155-83                 & F-L       & 1.865    & 2010/7/5-2010/7/17       & 44.85      & 46.44      & 45.23    & 47.86        & 45.19      & 1    \\
2FGL J2211.9+2355  & PKS 2209+236                & F-L       & 1.125    & 2009/4/15-2009/4/21      & 44.44      & 45.84      & 45.20    & 46.55        & 44.79      & 1    \\
2FGL J2225.6-0454  & 3C 446                      & F-L       & 1.404    & 2010/5/22-2010/5/27      & 45.70      & 46.55      & 45.68    & 47.41        & 45.60      & 1    \\
2FGLJ2253.9+1609   & 3C454.3                     & F-L       & 0.859    & 2009/12/4-2009/12/6      & 45.29      & 46.81      & 45.91    & 48.80        & 45.39      & 1    \\
2FGL J2258.0-2759  & PKS 2255-282                & F-L       & 0.926    & 2010/5/20-2010/5/26      & 44.87      & 46.42      & 45.25    & 47.24        & 45.84      & 1    \\
2FGL J2322.2+3206  & B2 2319+31                  & F-L       & 1.489    & 2009/5/20                & 44.41      & 45.75      & 45.08    & 47.11        & 44.71      & 1    \\
2FGL J2327.5+0940  & PKS 2325+093                & F-L       & 1.841    & 2010/6/18-2010/6/29      & 44.78      & 45.77      & 45.94    & 47.96        & 45.20      & 1    \\
2FGL J2334.3+0734  & TXS 2331+073                & F-L       & 0.401    & 2009/12/20               & 43.62      & 44.87      & 44.52    & 45.60        & 44.93      & 1    \\
2FGLJ2345.0-1553   & PMN 2345-1555/BZQJ2345-1555 & F-L       & 0.621    & 2009/1/10                & 43.70      & 45.83      & 44.25    & 46.54        & 44.36      & 1    \\
2FGL J2347.9-1629  & PKS 2345-16                 & F-L       & 0.576    & 2009/12/04-2009/12/05    & 44.39      & 45.73      & 45.01    & 46.24        & 44.36      & 1    \\
2FGLJ0854.8+2005   & OJ 287                      & B-L       & 0.3056   & 2010/4/26                & 43.86      & 46.03      & 44.85    & 45.91        & 43.205      & 2    \\
2FGLJ0958.6+6533   & TXS 0954+658                & B-L       & 0.368    & 2010/1/23                & 43.09      & 45.67      & 45.01    & 45.84        & 42.45      & 3    \\
2FGLJ1806.7+6948   & 3C 371                      & B-I       & 0.051    & 2009/11/3                & 41.73      & 44.54      & 42.71    & 44.16        & 42.00      & 2    \\
2FGLJ2202.8+4216   & BL Lac                      & B-L       & 0.0686   & 2009/11/11               & 42.69      & 45.16      & 44.08    & 45.05        & 42.52      & 3    \\
2FGLJ0238.7+1637   & AO 0235+164                 & B-L       & 0.94     & 2008/9/10                & 44.74      & 46.54      & 45.63    & 47.98        & 44.13      & 4    \\
2FGLJ0538.8-4405   & PKS 0537-441                & B-L       & 0.894    & 2008/10/9                & 44.85      & 47.12      & 45.70    & 47.87        & 45.31      & 4    \\
2FGLJ1420.2+5422   & 1418+546                    & B-I       & 0.152    & 2010/6/28                & 42.70      & 45.05      & 43.75    & 44.63        & 43.38      & 6    \\
2FGLJ1146.8-3812   & 1144-379                    & B-L       & 1.048    & 2010/6/24                & 44.94      & 46.13      & 45.54    & 46.80        & 44.97      & 6    \\
2FGLJ0523.0-3628   & PKS 0521-36                 & B-L       & 0.056546 & 2010/3/5                 & 42.50      & 44.53      & 43.66    & 44.48        & 42.68      & 3    \\
2FGLJ1800.5+7829   & S5 1803+784                 & B-L       & 0.684    & 2009/10/13               & 44.62      & 46.65      & 45.26    & 46.75        & 44.85      & 3    \\
2FGLJ1104.4+3812   & Mkn 421                     & B-H       & 0.030021 & 2009/11/15-2009/11/17    & 40.10      & 45.28      & 45.32    & 43.67        & 41.70      & 3    \\
2FGLJ1653.9+3945   & Mkn 501                     & B-H       & 0.033663 & 2010/3/21                & 41.47      & 44.31      & 44.39    & 43.24        & 42.20      & 3    \\
2FGLJ1517.7-2421   & 1514-241                    & B-L       & 0.049    & 2010/2/20                & 42.00      & 44.30      & 43.32    & 44.09        & 41.82      & 6    \\
2FGLJ0854.8+2005   & PKS 0851+202                & B-L       & 0.36     & 2010/4/10                & 44.02      & 46.02      & 44.64    & 45.68        & 43.74      & 6    \\
2FGLJ1221.4+2814   & Wcom                        & B-I       & 0.102    & 2010/02/17-2010/02/18    & 41.79      & 44.86      & 43.30    & 44.65        & 41.36      & 5    \\
\enddata
\end{deluxetable}

\clearpage
\begin{table}
	\centering
	\caption{The correlations of the whole blazar sample}
	\label{tab:Table 2}
	\begin{tabular}{lllllll} 
		\hline\hline
         &   & rad(5GHz)    & opt(5100 $\mathring{A}$)    & X(1keV)      & $\gamma$(100MeV)    & BLR    \\
         \hline
rad(5GHz)   & r      &      & 0.623     & 0.387     & 0.796     & 0.771     \\
         & P      &           & $1.411 \times 10^{-09}$ & $5.041 \times 10^{-04}$ & $5.463 \times 10^{-18}$ & $2.313 \times 10^{-16}$ \\
opt(5100 $\mathring{A}$)   & r      & 0.623     &      & 0.538     & 0.718     & 0.414     \\
         & P      & $1.411 \times 10^{-09}$ &           & $4.425 \times 10^{-07}$ & $2.057 \times 10^{-13}$ & $1. \times 10^{-04}$ \\
X(1keV)     & r      & 0.387     & 0.538     &      & 0.350     & 0.210     \\
         & P      & $5.041 \times 10^{-04}$ & $4.425 \times 10^{-07}$ &           & $1.821 \times 10^{-03}$ & $6.712 \times 10^{-02}$ \\
$\gamma$(100MeV) & r      & 0.796     & 0.718     & 0.350     &      & 0.673     \\
         & P      & $5.463 \times 10^{-18}$ & $2.057 \times 10^{-13}$ & $1.821 \times 10^{-03}$ &           & $2.000 \times 10^{-11}$ \\
BLR   & r      & 0.771     & 0.414     & 0.210     & 0.673     &      \\
         & P      & $2.313 \times 10^{-16}$ & $1.849 \times 10^{-04}$ & $6.712 \times 10^{-02}$ & $2.000 \times 10^{-11}$ &         
         \\
		\hline
	\end{tabular}
\end{table}

\begin{table}
	\centering
	\caption{The correlations of FSRQ sample}
	\label{tab:Table 3}
	\begin{tabular}{lllllll} 
		\hline\hline
         &   & rad(5GHz)    & opt(5100 $\mathring{A}$)    & X(1keV)      & $\gamma$(100MeV)    & BLR    \\
         \hline
rad(5GHz)   & r &      & 0.574     & 0.416     & 0.585     & 0.549     \\
         & P &           & $1.056 \times 10^{-06}$ & $7.646 \times 10^{-04}$ & $6.059 \times 10^{-07}$ & $3.887 \times 10^{-06}$ \\
opt(5100 $\mathring{A}$)   & r & 0.574     &      & 0.454     & 0.662     & 0.200     \\
         & P & $1.056 \times 10^{-06}$ &           & $2.084 \times 10^{-04}$ & $4.519 \times 10^{-09}$ & $1.188 \times 10^{-01}$ \\
X(1keV)     & r & 0.416     & 0.454     &      & 0.276     & 0.047     \\
         & P & $7.646 \times 10^{-04}$ & $2.084 \times 10^{-04}$ &           & $3.004 \times 10^{-02}$ & $7.179 \times 10^{-01}$ \\
$\gamma$(100MeV) & r & 0.585     & 0.662     & 0.276     &      & 0.373     \\
         & P & $6.059 \times 10^{-07}$ & $4.519 \times 10^{-09}$ & $3.004 \times 10^{-02}$ &           & $2.825 \times 10^{-03}$ \\
BLR   & r & 0.549     & 0.200     & 0.047     & 0.373     &      \\
         & P & $3.887 \times 10^{-06}$ & $1.188 \times 10^{-01}$ & $7.179 \times 10^{-01}$ & $2.825 \times 10^{-03}$ &  
      \\   
		\hline
	\end{tabular}
\end{table}

\begin{table}
	\centering
	\caption{The parameters of linear regression analysis for the whole blazar sample}
	\label{tab:Table 4}
	\begin{tabular}{llllll} 
		\hline\hline
                         & a    & b     & r(bisector) & r(Spearman) & p(Spearman) \\
                         \hline
$\log L_{BLR}$ VS $\log L_{rad}$ & 0.999 $\pm$0.064 & -0.256 $\pm$2.832 & 0.846      & 0.694      & $1.870 \times 10^{-12}$    \\
$\log L_{BLR}$ VS $\log L_{opt}$ & 0.663 $\pm$0.065 & 16.436 $\pm$2.882 & 0.621      & 0.500      & $3.203 \times 10^{-06}$    \\
$\log L_{BLR}$ VS $\log L_{X}$   & 0.889 $\pm$0.106 & 5.426 $\pm$4.718  & 0.436      & 0.541      & $3.089 \times 10^{-07}$    \\
$\log L_{BLR}$ VS $\log L_{\gamma}$   & 1.086 $\pm$0.076 & -1.574 $\pm$3.385 & 0.814      & 0.656      & $7.163 \times 10^{-11}$      
      \\   
		\hline
	\end{tabular}
\end{table}
\end{document}